 \documentclass[aps,prl,reprint,groupedaddress,showpacs]{revtex4-1}

\usepackage{graphicx,subfigure}
\usepackage{longtable}

\newcommand{\specialcell}[2][c]{%
  \begin{tabular}[#1]{@{}c@{}}#2\end{tabular}}

\usepackage[version=3]{mhchem} % Formula subscripts using \ce{}
\usepackage{graphicx,subfigure,amsfonts}

\begin{document}

%Title of paper
\title{A size-independent law to describe the alignment of shape-anisotropic objects}

\author{Ulla Vainio}
\email[]{ulla.vainio@aalto.fi}
%\homepage[]{Your web page}
%\thanks{}
%\altaffiliation{}
\affiliation{Institute of Materials Research, Helmholtz-Zentrum Geesthacht Centre for Materials and Coastal Research, Max-Planck-Str. 1, 21502 Geesthacht, Germany\\
Current address: Department of Applied Physics, Aalto University, Espoo, Finland}

%Collaboration name if desired (requires use of superscriptaddress
%option in \documentclass). \noaffiliation is required (may also be
%used with the \author command).
%\collaboration can be followed by \email, \homepage, \thanks as well.
%\collaboration{}
%\noaffiliation

\date{\today}

\begin{abstract}
A major challenge in the field of nanosciences is the assembly of anisotropic nano objects into aligned structures. The way the objects are aligned determines the physical properties of the final material. In this work, we take a closer look at the shapes of orientation distributions of aligned anisotropic nano and macro objects by examining previously published works. The data shows that the orientation distribution shape of anisotropic objects aligned by shearing and other commonly used methods varies size-independently between Laplace and Gaussian depending on the distribution width and on the cohesivity of the particles.
\end{abstract}

% insert suggested PACS numbers in braces on next line

%%%%%%%%%%%%%%%%%%%%%%%%%%%%%%%%%%%%%%%%%%%%%%%%%%%%%%%%%%%%%%%%%%%%%
%% Some journals require a list of abbreviations or keywords to be
%% supplied. These should be set up here, and will be printed after
%% the title and author information, if needed.
%%%%%%%%%%%%%%%%%%%%%%%%%%%%%%%%%%%%%%%%%%%%%%%%%%%%%%%%%%%%%%%%%%%%%

%\maketitle must follow title, authors, abstract, \pacs, and \keywords
\maketitle

\section{Introduction}
Production of aligned arrangements of anisotropic objects is a challenging task. While for example carbon nanotubes and silicon nanowires can be conveniently grown into arrays, the alignment of these structures is often far from perfect. \cite{Woodruff2007,Vainio2014} When oriented growth is not an option, anisotropic objects can be aligned by mechanical agitation such as shear \cite{Borzsonyi2012a,Borzsonyi2012b}, flow, \cite{Carastan2013} or vibration \cite{Yadav2013}, and sometimes a magnetic or electric field can act as the orientating agent. \cite{Song2013} Both in the macro scale and nano scale, most methods fail to give perfect alignment. The factors limiting the alignment are not fully understood.

In a composite material, the shape and width of the orientation distribution of its building blocks translates directly into other physical properties of the material. For example, the orientation distribution of carbon nanotubes within fibre ropes has a significant effect on the mechanical properties of the ropes. \cite{Liu2003} Theoretical studies show that the orientation distribution shape selected for simulation of a carbon nanotube film has a drastic effect on the electrical properties of the film. \cite{Simoneau2013} Nevertheless, instead of carefully examining the shape of the orientation spread, most experimental studies on alignment of nano and macro scale objects have been focused on obtaining a single number, an average alignment or an order parameter. In some cases, the orientation distributions of particle assemblies have been described with Gaussians \cite{Borzsonyi2012b,Boden2014}, Lorentzians \cite{Hwang2000}, the combination of both \cite{Wang2006,Das2011}, and even with squared Lorentzians  \cite{Trottier2008}. Typically, these functions did not fit the data perfectly but discrepancies between the data and the model were not discussed. Recently, a new better-fitting function, the generalized normal distribution, was applied to carbon nanotube orientation distributions which had been measured with high statistical accuracy using synchrotron radiation. \cite{Vainio2014} If this function could be applied also to other systems, it could be a game-changer for the study of aligned structures. In this contribution, we show that the generalized normal distribution fits to previously published data for many objects from nanometre to centimetre sizes, and we can now compare the shapes of the different orientation distributions to each other by using the same set of parameters.

\section{Experimental}
\begin{figure}
  \includegraphics[width=0.45\textwidth]{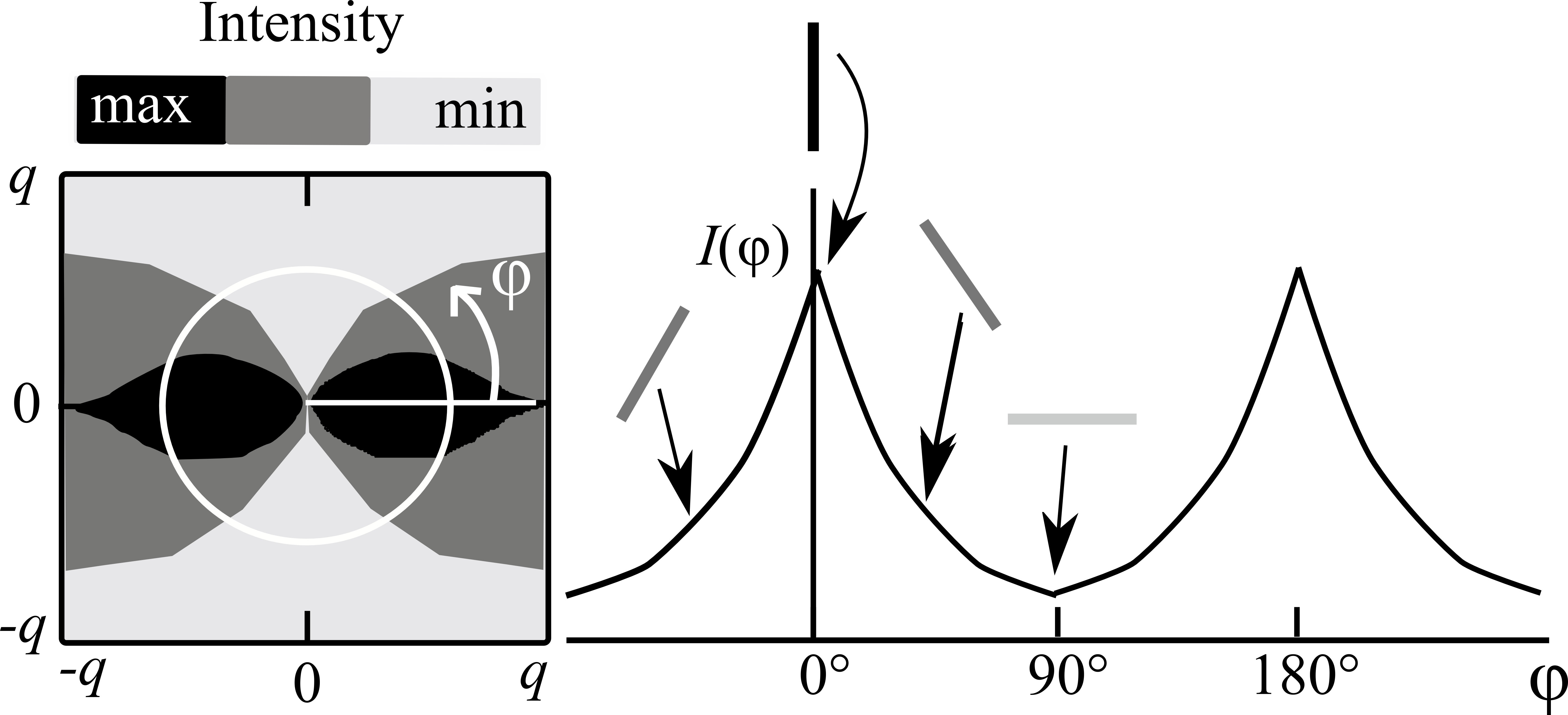}
  \caption{Schematic illustration of a small-angle scattering pattern from an oriented structure of rods, on the left, and the scattered intensity along the azimuthal angle $\varphi$ along the circle, on the right. The intensity at each $\varphi$ is related to the propability of finding a rod oriented in this angle within the sample.}
  \label{fig:azimuth}
\end{figure}

A survey of the literature shows that one of the most common ways to define the orientation of particle assemblies from X-ray scattering experiments or the like is by calculating the Hermans orientation parameter \cite{Wang2007,Hermans1946}
\begin{equation}
f = \frac{1}{2}\left(3\left<\cos^{2}\varphi\right> - 1\right),
\end{equation}
where the mean-square cosine is calculated from the scattered intensity $I(\varphi)$ by integrating over the azimuthal angle $\varphi$ (see Fig.~\ref{fig:azimuth})
\begin{equation}
\left<\cos^{2}\varphi\right> = \frac{\int_{0}^{\pi/2}I(\varphi)\sin\varphi\cos^{2}\varphi d\varphi}{\int_{0}^{\pi/2}I(\varphi)\sin\varphi d\varphi}.
\end{equation}
For perfect vertical orientation $f$ = 1, for isotropic orientation $f$ = 0, and for perfect horizontal orientation $f$ = -0.5. In the following analysis, we have calculated the orientation parameter for all distributions as if they were perfectly vertically centred by shifting the $\varphi = 0^\circ$ accordingly in order to compare the orientation degree rather than direction of alignment. The orientation parameter can be calculated for any orientation distribution regardless of their shape.

In the next step, we compare the values of Hermans orientation parameter to parameters describing the shape of the orientation distribution for different experimentally observed particle assembly systems by refitting literature data. Recently, it was identified that the shape of the orientation distribution of multiwalled carbon nanotube arrays (MWCNTs) can be modelled accurately with a family of symmetric distributions that include all shapes between Laplace and Gaussian. \cite{Vainio2014} This {\it generalized normal distribution} (GND), also called {\it exponential power distribution} in the literature \cite{Nadarajah2005,Subbotin1923,Nadarajah2006} has the form
\begin{equation}
p_{0}(\varphi) = \frac{\beta}{2\alpha\Gamma(1/\beta)}\exp\left[-\left(\frac{|\varphi - \mu|}{\alpha}\right)^\beta\right],
\label{gnd}
\end{equation}
where $\alpha$ is a scaling factor related to the width, $\beta$ is the shape parameter determining the sharpness, and $\mu$ is the mean of the distribution. $\Gamma$ denotes the gamma function. The GND reduces to the normal distribution when $\beta$ = 2 and to Laplace distribution when $\beta$ = 1. Due to its generality, this distribution has actually been invented several times in the course of history. In diffusion studies it is known as the {\it stretched exponential function} with $0< \beta < 1$, and in the field of relaxation dynamics in materials, it is also called the {\it Kohlrausch function} [or {\it Kohlrausch-Williams-Watts function} (KWW)] after the physicist Rudolf Kohlrausch who applied it in the 19th century to describe electric charge decay.  \cite{Kohlrausch1854,Cardona2007}

\begin{figure*}
	\subfigure[]{%
  		\includegraphics[width=0.45\textwidth]{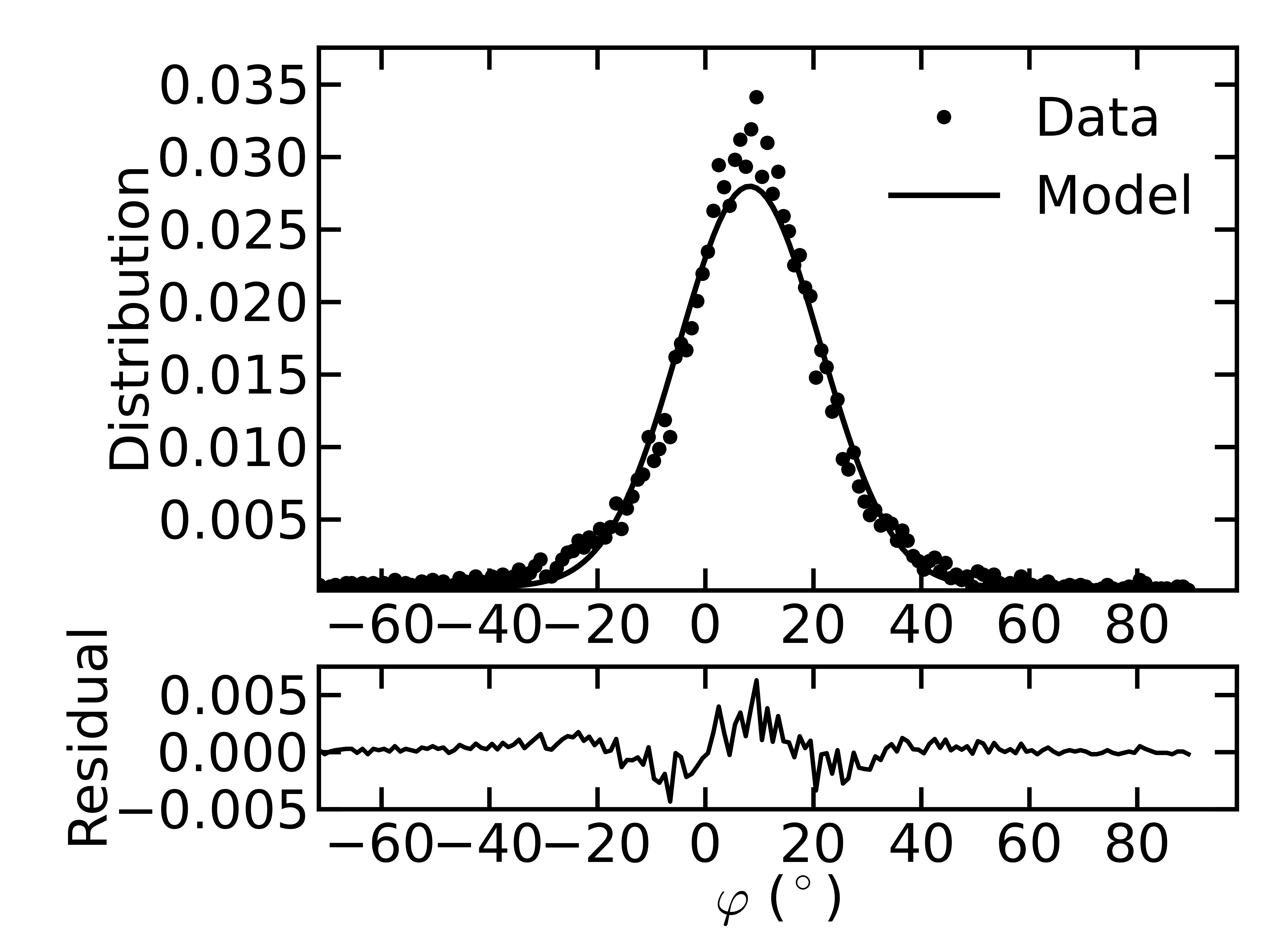}
	}%
	\subfigure[]{%
		  \includegraphics[width=0.45\textwidth]{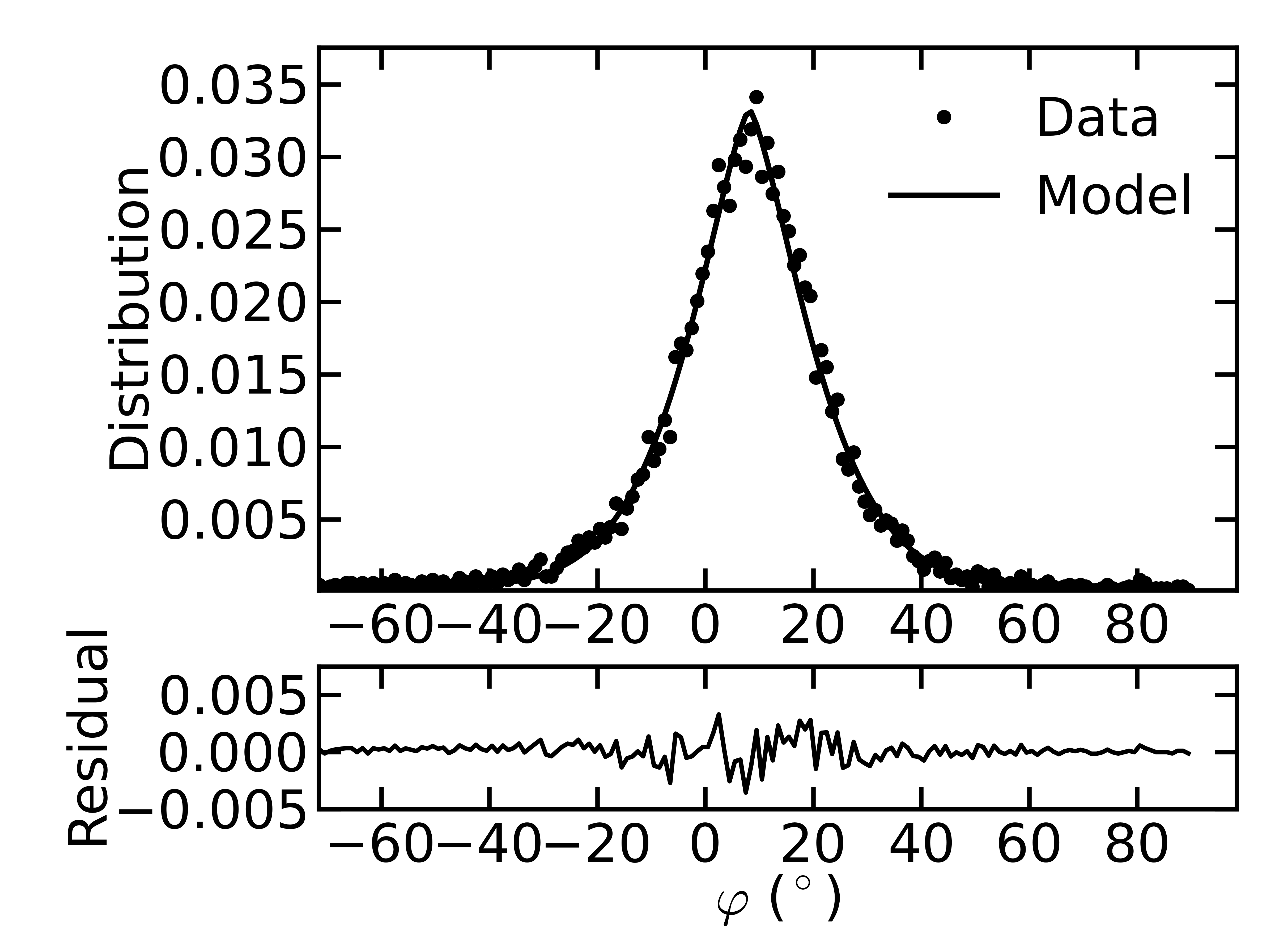}
	}%
  \caption{Orientation distribution of wooden pegs in a rheometer \cite{Borzsonyi2012b} fitted with (a) a Gaussian and (b) a generalized normal distribution.}
  \label{fig:fitcomparison}
\end{figure*}

In most cases in the literature, uncertainties for the data were not available so goodness-of-fit is not reported here, but the relative likelihoods for model selection were calculated from the residual sum of squares RSS = $\sum_{n=1}^N (y_n - p(\varphi_n))^2$. We use the Akaike information criterion,\cite{Akaike1974} AIC = $2k - N \ln(\textrm{RSS})$, to compare the relative likelihoods, $\exp((\textrm{AIC}_{\textrm{min}}-\textrm{AIC})/2)$, of the Gaussian, Lorentzian and generalized normal distribution for the model selection. Here, $k$ is the number of free parameters, $N$ the number of data points, $p(\varphi)$ is one of the three models, and AIC$_{\textrm{min}}$ is the smallest AIC value obtained for the models. Fig.~\ref{fig:fitcomparison} depicts one example of of a fit with Gaussian and GND. 

It should be noted that the measured orientation distribution is often merely a projection of the real one. Methods which rely on counting the angles of individual particles in two-dimensional images cannot capture the shape of the three-dimensional orientation distribution. But since this is the case also for many scattering methods, such as small-angle X-ray scattering, where we also see only the projection of the three-dimensional distribution, the experimental orientation distributions should be comparable to each other. The effect of projection on the shape of the orientation distribution is not dramatic: The projection of a normal distribution is a normal distribution, and the projection of a generalized normal distribution is a generalized normal distribution, only with different $\beta$ and $\alpha$ because the projection shifts the shape of the distribution closer to a Gaussian. \cite{Vainio2014}

Simulations of two-dimensional granular structures of cohesive, elongated particle assemblies can be found in the literature. \cite{Hidalgo2012} In one of the the simulations, elongated particles with aspect ratio of 10 are dropped with random orientations onto a pile and the sticking coefficient of the particles is varied. The reported orientation distributions from this study are not exactly equivalent to the projected orientation distribution observed in many experiments, because the orientation of particles is defined solely with respect to horizontal plane, but the generalized normal distribution fits the simulated data: for non-sticking particles (Bond number  Bog = 0, Simulation1), slightly sticking particles (Bond number  Bog = $10^3$, Simulation2), and strongly sticking particles (Bond number  Bog = $10^4$, Simulation3) shape factors turn out to be $\beta$=1.48, 1.36, and 1.62, respectively.

\section{Results}

%\begin{turnpage}
\begin{table*}
\caption{\label{tab:Shape}Descriptions of objects and shape, $\beta$, and scale, $\alpha$, parameters of their orientation distributions according to a fit with the generalized normal distribution. $D$ denotes the diameter of the object and $L/D$ its aspect ratio. For MWCNTs and Al$_2$O$_3$ platelets the range of values show the variation within one sample. Bad fit with the generalized normal distribution is marked with a dash (-) for $\alpha$ and $\beta$.}
		\begin{tabular}{lllllllll}
			Objects & D ($\mu$m) & $L/D$ & Alignment & $\beta$ & $\alpha$ ($^\circ$) & Method & Ref \\
			\hline
			CdSe nanorods &  0.008 & 2.75 & rubbing & 1.83 & 31 &GIWAXS & \cite{Breiby2009} \\
			Polymer cryst. &  $< 0.017$ & -  & strain & - &  - &  XRD & \cite{Kamal2012} \\
			MWCNTs &  0.040--0.070 & -  & grown  & 1.37--1.65 &  26--38 & SAXS & \cite{Vainio2014} \\
			Cellulose whiskers 1 &  1.95 & 4.1 & magnetic  & 1.35 & 12 & XRD & \cite{Song2013} \\
			Cellulose whiskers 2 &  7.18 & 3.2 & magnetic  & 1.19 & 21 & XRD & \cite{Song2013} \\
			Al$_2$O$_3$ platelets 1 &  10 & 0.03--0.05  &  sediment  & 1.21--1.55 & 19-28 & XRD & \cite{Behr2014}\\
			Al$_2$O$_3$ platelets 2 &  10 & 0.03--0.05  & pressed & 1.55$\pm$0.06 & 17.4$\pm$0.4 & XRD & \cite{Behr2014}\\
			Cellulose whiskers 3 &  16.1 & 6.4  & magnetic  & 2.07 & 42 & XRD & \cite{Song2013} \\
			Rice 1 & 1600 & 4.5 & shear & 1.59 & 20 & optical & \cite{Borzsonyi2012b} \\
			Glass cylinders &  1900 & 3.5 & shear  & 1.63 & 27 & optical & \cite{Borzsonyi2012b} \\
			Rice 2 & 2000 & 3.4 & shear  & 1.63 & 23 & optical & \cite{Borzsonyi2012b} \\
			Rice 3 & 2800 & 2.0 & shear  & 1.69 & 28 & optical & \cite{Borzsonyi2012b} \\
			Wooden pegs &  $>5000$ & 5.0 & shear & 1.39 & 15 & X-ray CT & \cite{Borzsonyi2012b} \\
			Simulation 1 &  - & 10 & \specialcell{gravitation\\($Bo_g$=0)} & 1.48 & 18 &\specialcell{2d\\simulation} & \cite{Hidalgo2012} \\
			Simulation 2 &  - & 10 & \specialcell{gravitation\\($Bo_g$=10$^3$)} &  1.36 & 20 &\specialcell{2d\\simulation} & \cite{Hidalgo2012} \\
			Simulation 3 &  - & 10 & \specialcell{gravitation\\($Bo_g$=10$^4$)}  & 1.62 & 35 &\specialcell{2d\\simulation} & \cite{Hidalgo2012} \\			
		\end{tabular}
\end{table*}
%\end{turnpage}

\begin{table}
\caption{\label{tab:AIK} Relative likelihoods of the Gaussian, RL(G), Lorentzian RL(L), and the generalized normal distribution, RL(GND), fitting the data have been calculated using the Akaike information criterion.}
		\begin{tabular}{lllll}
			Objects & RL(G) & RL(L) & RL(GND) \\
			\hline
			CdSe nanorods & 1.0 & 10$^{-50}$ & 10$^{-27}$  \\
			Polymer cryst. & - & - & - \\
			MWCNTs  & 10$^{-89}$& 10$^{-28}$ & 1.0  \\
			Cellulose whiskers 1 & 10$^{-37}$ & 10$^{-17}$ & 1.0  \\
			Cellulose whiskers 2 & 10$^{-31}$ & 0.001 & 1.0   \\
			Al$_2$O$_3$ platelets 1 & 10$^{-22}$ & 1.0 & 10$^{-4}$ \\
			Al$_2$O$_3$ platelets 2  & 10$^{-7}$ & 10$^{-38}$ & 1.0 \\
			Cellulose whiskers 3  & 1.0 & 0.001 & 0.486  \\
			Rice 1  & 10$^{-32}$ & 10$^{-85}$ & 1.0  \\
			Glass cylinders & 10$^{-22}$ & 10$^{-42}$ & 1.0  \\
			Rice 2 & 10$^{-33}$ & 10$^{-78}$ & 1.0  \\
			Rice 3  & 10$^{-19}$ & 10$^{-58}$ & 1.0  \\
			Wooden pegs  & 10$^{-25}$ & 10$^{-35}$  & 1.0  \\
			Simulation 1 & 10$^{-39}$ & 10$^{-49}$ & 1.0  \\
			Simulation 2  & 10$^{-29}$ & 10$^{-13}$ & 1.0  \\
			Simulation 3  & 10$^{-10}$ & 1.0 & 0.016  \\			
		\end{tabular}
\end{table}

Table~\ref{tab:Shape} summarizes fit results of the generalized normal distribution to published literature data. For the majority of cases, the GND gave the best fit (Table~\ref{tab:AIK}), but close to $\beta = 2$ and for large $\alpha$ the Gaussian is better which is reasonable as the Gaussian model has one free parameter less than the GND model. For two cases the Lorentzian fits the best, with GND being the second best model.

It should be noted that the alignment spread in the case of cellulose whiskers \cite{Song2013} might be due to orientation spread of cellulose crystallites within the whiskers rather than due to incomplete alignment in magnetic field.

\begin{figure}
  \includegraphics[width=0.5\textwidth]{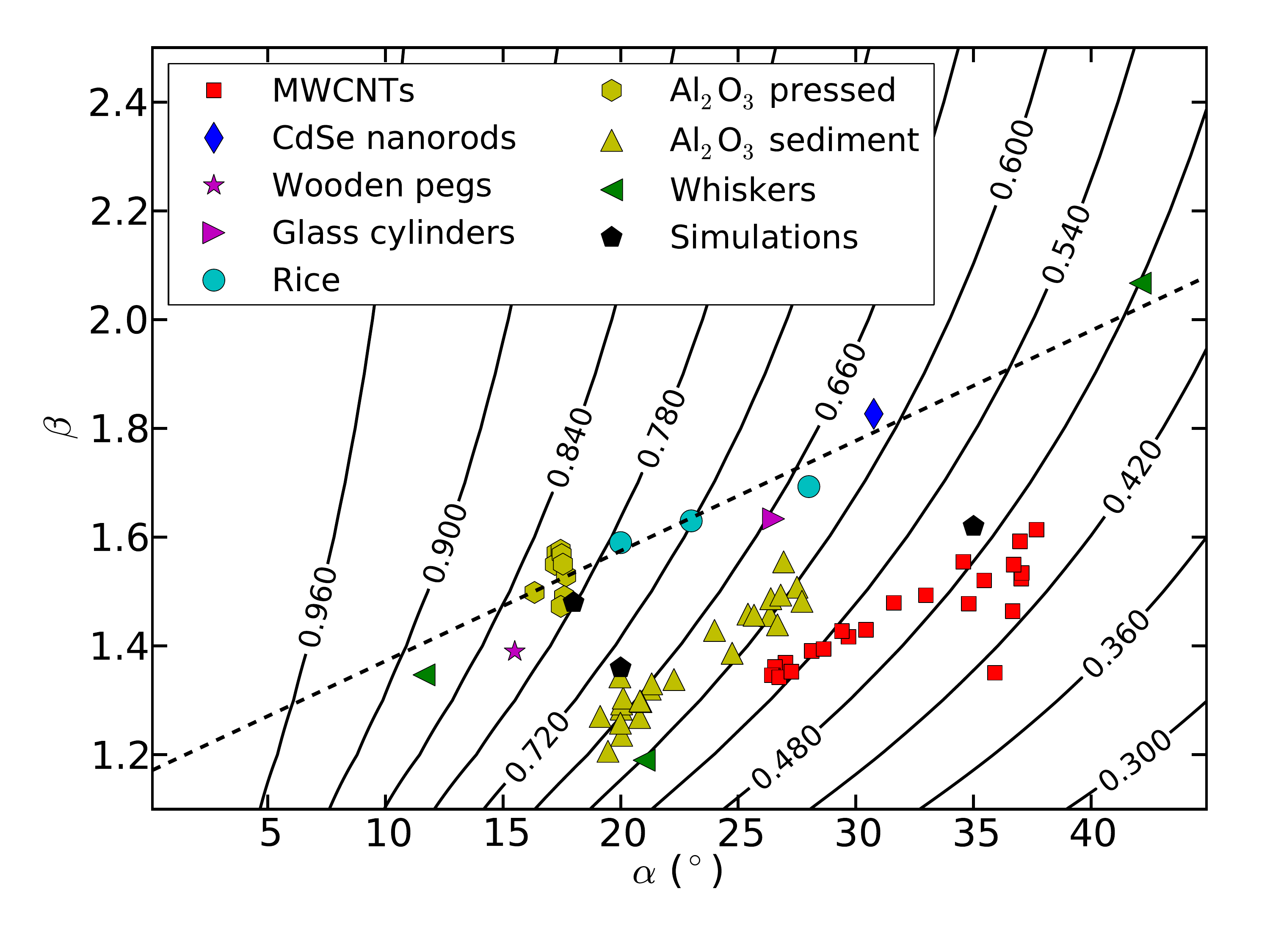}
  \caption{(color online) Orientation distribution shape, $\beta$, marked with symbols, as a function of scale parameter $\alpha$. The values of Hermans orientation parameter are marked with contour lines. All parameters were obtained from fits with the generalized normal distribution (equation (\ref{gnd})) to experimental data (table~\ref{tab:Shape}). The dashed line ($\beta(\alpha) = 0.020\alpha + 1.17$) is a fit to the data from rice and CdSe nanorods. For multiwalled carbon nanotubes and Al$_2$O$_3$ platelets each data point represent a different position on one sample.}
  \label{fig:hermans}
\end{figure}

Fig.~\ref{fig:hermans} visualises the data presented in Table~\ref{tab:Shape}. The contour lines mark the Hermans orientation parameter and this shows that the best orientation is found for high $\beta$ and low $\alpha$ values, in the upper left corner of the graph. All of the experimental data is located in the lower right corner of the graph and there seems to be no prominent difference between nano and macro scale particles. The dashed line showing a fit to a few data points in which alignment has been obtained by shearing shows a sort of a limit to the orientation distribution shape. It is possible to achieve near to perfect alignment using these commonly used methods for alignment but then, according to this graph, we should expect the orientation distribution to be more Laplace like than Gaussian.

In the case of Al$_2$O$_3$ platelets intercalated with polymer, two data sets for same particle type are available. The first one of sedimented particles shows poorer alignment than a second set where the sedimented particle assembly was further compressed. The orientations of sedimented Al$_2$O$_3$ platelets should be dominated more by cohesive forces than the sedimented and pressed platelets. Effect of a moderate amount of cohesion is seen both in simulation \cite{Hidalgo2012} and experiment as a decrease in $\beta$ and increase in $\alpha$. The simulation with most sticky particles results in a broad orientation distribution with increased $\beta$. The carbon nanotube forests could be described effectively as very sticky granular systems. The most prominent result from cohesion is the decrease in the overal orientation degree of the particles.

Despite the success in fitting most of the data presented in table~\ref{tab:Shape}, there is also one data set which could not be fitted with the GND. In case of polymer crystallites of poly($\varepsilon$-caprolactone) oriented under strain, \cite{Kamal2012} the orientation of crystallites did not follow the generalized normal distribution. This is an example of a system which is composed of particles that are interconnected. This situation is very different from all the other cases presented here. The applicability of the generalized normal distribution may very well be limited only to particle assemblies which allow free movement of the particles.

\section{Discussion}
Now that we have identified the generalized normal distribution to be a feasible model for a multitude of particulate systems, we need to consider its physical meaning. There exist several theoretical models for orientation distributions of particles in different environments, and it is not clear if some of them could actually have the same shape as the generalized normal distribution. Fitting a non-cyclic function to the orientation distribution as a function of azimuthal angle $\varphi$ is not fully correct, because it cannot describe all the situations between isotropically oriented and fully oriented systems. A mathematically correct model would need to have cyclic properties. Next, we inspect cyclic functions found in the literature to see if they could actually reproduce the shape of the generalized normal distribution. Theoretical framework for the orientation distribution shape exists for example in the case of spheroidal particles in dilute suspension under shear. The function describing the orientation of spheroids of aspect ratio $r_e$ is given \cite{Mueller2010}
\begin{equation}
p_1(\varphi) \propto \frac{1}{r_e^2\cos^2\varphi + \sin^2\varphi},
\end{equation}
where $\varphi$ is the misorientation of the symmetry axis of the particle compared to the flow direction. This shape should be valid also for other centrosymmetric particles, such as cylinders and discs, but fitting this function with the generalized normal distribution did not produce satisfying results.

The Maier-Saupe distribution, which can be applied to describe the orientation distributions in liquid crystals and to study the chain orientation in cholesterol-lipid systems, \cite{Mills2008} is a much more promising candidate for the physical background of the generalized normal distribution. A thorough examination of the Maier-Saupe distribution for scattering data is given by Mills et al. \cite{Mills2008}, and they define it
\begin{equation}
p_2(\varphi) = \exp\left(\frac{m\cos^2\varphi}{2}\right)I_0\left(\frac{m\cos^2\varphi}{2}\right),
\label{maiersaupe}
\end{equation}
where $I_0$ is a modified Bessel function of the first kind and $m$ is a parameter related to the width of the distribution. For simplicity, we have omitted the normalization factor in equation~(\ref{maiersaupe}) but it can be found in the original publication. While $p_2$ can be fitted to great accuracy (but not perfectly) with the generalized normal distribution, the shape factor, $\beta$, remains above 1.73 for all parameter values of the Maier-Saupe distribution and hence the Maier-Saupe model cannot be the correct model to use in the case of most of the systems presented here.

A special orientation distribution has been used to simulate acoustic  non-woven fibre systems consisting of cylindrical subunits. \cite{Schladitz2006} This distribution is characterized mainly by the anisotropy parameter $p$:
\begin{equation}
p_3(\theta,\varphi) = \frac{p \sin \theta}{4 \pi \left[1 + (p^2-1)\cos^2\theta\right]^{3/2}}.
\end{equation}
Here, $\theta \in [0,\pi)$ and $\varphi \in [0,2\pi)$ are the altitude and longitude in spherical coordinates. The $\sin \theta$ term in this equation is responsible for assigning the correct propability to each $\theta$ when we are interested in the volume orientation distribution but we may compare the number orientation distributions to each other without this normalization such that $p_3(\theta,\varphi)/\sin \theta$ is constant for $p = 1$. For $p < 1$, the cylinders are more oriented along the symmetry axis. Again, $p_3(\theta,\varphi)/\sin \theta$ does not have the shape of the generalized normal distribution.

For modelling of orientations of graphene layers, the projected von Mises-Fisher distribution has been introduced \cite{Boehlke2013}
\begin{equation}
p_4(\varphi,\gamma) = \frac{\kappa}{4\sinh \kappa}L_{-1}(\kappa \cos (\gamma - \varphi)).
\end{equation}
Here $\varphi \in [0,\pi]$ is the azimuthal angle, $\gamma \in [0,\pi]$ represents the angle of preferred orientation, $\kappa$ is a concentration parameter, and $L_{-1}$ is the modified Struve function. The von Mises-Fisher distribution is a directional analogue of the Gaussian distribution and hence it cannot reproduce the shapes of the generalized normal distribution, apart from $\beta$ = 2.

None of the cyclic functions, $p_1 - p_4$, presented above are able to capture the range of distribution shapes which we found in real systems. In conclusion, despite the shortcomings due to non-cyclicity, the generalized normal distribution is at the moment the most suited function for the study of moderately aligned systems, even if it cannot be used to describe systems close to isotropic alignment.

\section{Conclusions}
Here we have shown that alignment of freely moving anisotropic objects both in nano and macro scale can be described by one function, the generalized normal distribution. Spread of the  experimental data in Fig.~\ref{fig:hermans} allows us to draw some general conclusions about the alignment of anisotropic objects. We observe that the projection of orientation distribution of anisotropic particles is close to the Laplace distribution ($e^{-|x|}$) when very good alignment is achieved. Exponential decay occurs commonly in the field physics, and the Laplacian orientation distribution may be a manifestation of an underlying relaxation processes, which follow an exponential decay. Moderate or poor alignment will lead to a more Gaussian distribution ($e^{-x^2}$) but slightly cohesive particles may behave differently. These findings should be taken into account in future studies of materials consisting of aligned anisotropic particles.

%%%%%%%%%%%%%%%%%%%%%%%%%%%%%%%%%%%%%%%%%%%%%%%%%%%%%%%%%%%%%%%%%%%%%
%% The "Acknowledgement" section can be given in all manuscript
%% classes.  This should be given within the "acknowledgement"
%% environment, which will make the correct section or running title.
%%%%%%%%%%%%%%%%%%%%%%%%%%%%%%%%%%%%%%%%%%%%%%%%%%%%%%%%%%%%%%%%%%%%%
\begin{acknowledgements}

I gratefully acknowledge financial support from the German Research Foundation (DFG) via SFB 986 "M$^3$", project Z2. T. B\"orzs\"onyi, D. Breiby, R. C. Hidalgo, T. Kamal, and T. Kimura are thanked for giving access to the original data from their publications. I. Krasnov, R. Gehrke and U. Handge are thanked for inspirational discussions. I thank Martin M\"uller and Andreas Schreyer for providing the perfect working conditions.

\end{acknowledgements}

%%%%%%%%%%%%%%%%%%%%%%%%%%%%%%%%%%%%%%%%%%%%%%%%%%%%%%%%%%%%%%%%%%%%%
%% The appropriate \bibliography command should be placed here.
%% Notice that the class file automatically sets \bibliographystyle
%% and also names the section correctly.
%%%%%%%%%%%%%%%%%%%%%%%%%%%%%%%%%%%%%%%%%%%%%%%%%%%%%%%%%%%%%%%%%%%%%
\providecommand{\latin}[1]{#1}
\providecommand*\mcitethebibliography{\thebibliography}
\csname @ifundefined\endcsname{endmcitethebibliography}
  {\let\endmcitethebibliography\endthebibliography}{}

\end{document}